\documentstyle[preprint,prl,aps]{revtex} 
\tightenlines
\begin{document}


\title{\bf Van der Waals contribution to 
the inelastic atom-surface scattering}
\author{M. Machado$^{1}$ and
D. S\'anchez-Portal$^{2}$}
\address{$^1$ Departamento de F\'{\i}sica de Materiales, Facultad de Qu\'{\i}mica, UPV-EHU \\
Apdo. 1072, 20080 Donostia, Spain\\
$^2$ Centro Mixto CSIC-UPV-EHU and Donostia International Physics Center (DIPC)\\
Paseo Manuel de Lardizabal, 4, 20018 Donostia, Spain\\}

\maketitle
\footnotetext[1]{ Electronic address: wabmagam@sc.ehu.es; Fax:
+34-943015600}

\begin{abstract}
A calculation of the inelastic scattering rate 
of Xe atoms on Cu(111) is presented. We focus in 
the regimes of low and intermediate 
velocities, where the energy loss is
mainly associated to the excitation electron-hole pairs
in the substrate. 
We consider trajectories parallel to the 
surface and restrict ourselves to
the Van der Waals contribution.
The decay rate is calculated within
a self-energy formulation.
The effect of the response
function of the substrate
is studied by comparing the results obtained
with two different approaches:
the Specular Reflection Model and the
Random Phase Approximation. In the latter, the surface
is described by a finite slab and the wave functions
are obtained from a one-dimensional
model potential that describes the main features
of the surface electronic structure while correctly retains
the image-like asymptotic
behaviour.
We have also studied the influence of the surface state on the
calculation, finding that it represents around $50\%$
of the total probability of electron-hole pairs excitation.

\end{abstract}
\newpage

An incident particle scattering with a solid metal 
surface losses energy mainly through
two channels: excitation of phonons and electronic 
excitations (electron-hole pairs,
plasmons, etc). At low energies, creation of 
phonons and electron-hole pairs
becomes the main available channel. 
Understanding the different mechanisms for energy 
loss is a difficult task that has
attracted both experimental and theoretical work. 
In an early experiment,
Amirav and Cardillo \cite{Cardillo} reported 
a method for measuring directly the 
electron-hole (e-h) pair excitation in the scattering 
of Xe atoms on a semiconducting surface.  
More recently, it has been possible to measure the 
e-h pairs created by adsorption 
of hydrogen and deuterium\cite{Nienhaus,Gergen,Nienhaus2}
on transition metal
surfaces.

In the theoretical side, e-h pair creation 
probabilities due to moving atoms have been calculated
with the use of several models. The most simple 
calculations were carried out within the
Specular Reflection Model (SRM)\cite{Ritchie}, 
in which the surface potential is replaced
by an infinite barrier\cite{Persson3}. This implies 
that the electronic density vanishes at
the surface,  
so the surface polarizability is not well described 
and energy losses are underestimated. More realistic 
calculations were performed
by using the jellium 
model\cite{Feibelman,Persson1,Persson2,Eguiluz,Liebsch}
but surface states, which represent an 
important channel of electron-hole pair creation,
are not included in this description.
Very few
calculations of the energy loss 
to date incorporate most of the complexity 
of the surface. One example, however, 
is the very recent ab initio calculation
of the
e-h pair creation by hydrogen scattering on Cu(111) 
performed by Trail et al.\cite{Trail}.
There is also a large amount of theoretical calculations 
of energy loss of charged particles, ranging from very simplified models for the response
function\cite{Echenique1,Echenique2,Zabala,Juaristi,Nagatomi}
to more accurate self-consistent jellium calculations\cite{Cazalilla,Lekue}. 
 
In this paper, we calculate the contribution 
of the long-range van der Waals term to
the probability of excitation of e-h 
pairs by an incident atom interacting with
a metal surface. This problem
was also studied by Annett and Echenique\cite{Annett}, in an 
attempt to explain
the experimental results of Amirav and Cardillo. 
They used a self-energy formalism, with
the response function described within the SRM, and 
obtained a probability four orders of
magnitude less than the experimental one. 
Several reasons might explain this large difference. 
On the
one hand, the use of the simple SRM model 
for the response function. On the other hand,
several other contributions to the scattering process 
were not taken into account, like those related to
the atom-surface overlap (shorter-range effects), 
interaction between the static dipole generated in the atom 
by the surface
proximity and the substrate, and non-adiabatic effects. 
We also focus on the van der Waals term, leaving the exploration
of the other terms to a future publication, 
and try to study the influence of the model used for
the response function, 
the semiclassical SRM or the more realistic Random 
Phase Approximation (RPA). RPA provides
a more correct description of the electron 
density at the surface region 
and permits to investigate the importance
of the surface state band in the electron-hole pair excitation.

Unless otherwise is stated, atomic units are used throughout, i.e., m=$e^2$=$\hbar$=1


We consider neutral atoms moving parallel 
to a metal surface with a constant, 
non relativistic velocity $v$, at a 
distance $z_0$ from the substrate, which occupies
the $z<0$ half-space.
We will start from the
Born-Oppenheimer wave function for the atom, 
$|\alpha>=\phi_{\alpha}({\bf R})
\psi_{\alpha}({\bf r}_1,{\bf r}_2,...;{\bf R})$,
where the ${\bf r}_i$ are the electronic 
coordinates and {\bf R} the nuclear ones.
Ground state wave functions
of the solid will be denoted by $|\beta>$, 
and $|\beta'>$ will be the excited states.
Neglecting coupling between the atom
and the substrate, the many-body wave function 
of the system can be written as the product 
of the respective wave functions of the atom and the solid.

The contribution to the energy loss
due to the long-range van der Waals coupling 
between the surface and the atom can be
obtained by defining a self-energy, whose real part 
represents the van der Waals energy, and its imaginary part is
related with the inelastic scattering rate, $\Gamma$, so that
$\Gamma=-2Im(\Sigma)$. The transition rate of the incident atom by
excitation of electrons from the Fermi sea to 
unoccupied levels can be calculated by
introducing the Coulomb operator, 
$\hat{V}$, as the coupling between atom 
and substrate in Fermi's golden rule. 
Therefore, the self-energy will be given by:

\begin{equation}\label{self2}
\Sigma_{\alpha\beta}=-\sum_{\alpha',\beta'}
\frac{|<\alpha\beta|\hat{V}|\alpha'\beta'>|^2}
{\epsilon_{\alpha'}+
\epsilon_{\beta'}-\epsilon_{\alpha}-\epsilon_{\beta}-i\delta}
\end{equation}

Where the $\epsilon$ are the corresponding 
energies of the states of the
incoming atoms and the substrate.
We can simplify eq.(\ref{self2}) by introducing 
the response function of the metal and the two-dimensional
Fourier transform of 
the Coulomb potential, obtaining:

\begin{eqnarray}\label{self4}
\Sigma_{\alpha}=-\sum_{\alpha'}
\int_0^{\infty}\frac{d\omega}{\pi}\frac{1}
{\omega+\epsilon_{\alpha'}-\epsilon_{\alpha}-i\delta}
\int d{\bf r_{\|}}  d{\bf r''_{\|}} 
dz dz' dz'' dz''' \frac{d^2{\bf q_{\|}}}{(2\pi)^2}
\left(\frac{2\pi}{q_{\|}}\right)^2 \\ \nonumber 
Im[\chi(q_{\|},\omega;z',z''')]
e^{i{\bf q}_{\|}({\bf r}_{\|}-{\bf r''}_{\|})}
e^{-q_{\|}|z-z'|}e^{-q_{\|}|z''-z'''|}
<\alpha|\hat{\rho}_a({\bf r})|\alpha'>
<\alpha'|\hat{\rho}_a^\dagger({\bf r''})|\alpha>
\end{eqnarray}

Where $\omega$ is the transferred energy, and 
$\hat{\rho}_a({\bf r})$ is the charge-density operator
for the atom.
Eq.(\ref{self4}) is valid for arbitrary trajectories. 
Now we focus on trajectories parallel
to the surface, replace the density operator by 
$\hat{\rho}_a({\bf r})=
Z\delta({\bf r}-{\bf R})-\sum_i\delta({\bf r}-{\bf r}_i)$, 
where Z is the
nuclear charge,
and
take nuclear wavefunctions as, 

\begin{equation}
\phi_{\alpha}({\bf R})=\frac{1}{(2\pi)^2} 
e^{(iM{\bf v_{\|}r_{\|}})}\delta(z-z_0)
\end{equation} 

Assuming the atom is far from the surface, we can
make a multipole expansion in which we 
take the dipole term as the leading contribution,
because for neutral atoms there is no monopole contribution. 
This term introduces the dipole 
oscillator strengths, corresponding 
to transitions from electronic state 0 to n, 
$f_{n0}=2\omega_{n0}|<n|{\bf r}|0>|^2/3$
where $\omega_{n0}$ are the frequencies of the dipolar transitions.
These transformations in eq.(\ref{self4}) 
lead to the following expression of the scattering rate: 

\begin{eqnarray}\label{prob}
\Gamma=-\sum_{n}\frac{f_{n0}}{\omega_{n0}}
\int_0^{\infty} d\omega \int d^2{\bf q}_{\|} 
\int dz'\int dz''' Im\chi(q_\|,\omega;z',z''')
e^{-q_\||z_0-z'|}e^{-q_\||z_0-z'''|}\\ \nonumber
(1+sig(z_0-z')sig(z_0-z''')) 
\delta(\omega+\omega_{n0}-{\bf{q_\|.v_\|}}+q^2/2M)
\end{eqnarray}

Where sig(z-z') stands for the sign function.
Once the energy transfer is fixed by Dirac's delta, 
maximum and minimum momentum transfers are given
by $q^{\pm}_{\|}=M(vcos\theta\pm\sqrt{v^2cos^2\theta-2\omega_{n0}/M})$. $\theta$ is the angle
between the parallel momentum and the initial velocity. 
M is the mass of the atom,
which for Xe atoms is of the order of $10^4$ a.u.
Energies above the 9 eV threshold, corresponding 
to the first available dipole transition in
the Xe atom, are needed so that e-h pair 
excitation is feasible. Therefore, it is the gap in the
atomic levels which determines the need 
of a minimum momentum transfer different from zero. $q^+$ 
corresponds to the transfer of the whole initial energy.

We have used two different models for the response function. 
In the SRM the surface potential is 
replaced by an infinite barrier and it is assumed that the electrons are specularly reflected at 
the surface, with no interference terms. 
Therefore, the surface response function is strictly zero
beyond the jellium edge. In the 
more realistic RPA model\cite{Lindhard,Review,Eguiluz1}, electrons
respond as a non-interacting gas to the total 
potential, and the response function is obtained
solving a self-consistent equation on the induced potential. 
We model the metal
by a slab, such that one-particle wave functions in the directions 
parallel to the surface are plane waves, while 
in the perpendicular direction they are the solutions of
a one-dimensional model
potential\cite{Chulkov}. 
This potential reproduces the width and position of the band gap, 
as well as the binding energies of
the surface and image states. Close to the Fermi level, the resulting
wave functions 
are comparable to those obtained with an ab initio calculation\cite{Review}. The potential has also been proved
to accurately predict the broadening of surface and image states on metal 
surfaces\cite{Chulkov2,Chulkov3,Chulkov4,Chulkov5}.

 
For the RPA calculations we have used a supercell 
geometry where the Cu(111) 
surface is represented by a 30-layers slab, 
and the vacuum region is equivalent to 20 atomic layers.
In the SRM model\cite{Ritchie},
the surface response is directly obtained 
from Mermin's dielectric function\cite{Mermin}.
We have used the 
values 2.67 and 0.013 a.u.for $r_s$ and the
damping parameter $\gamma$ respectively. Our results exhibit only
a slight dependence on $\gamma$. This is a clear indication that,
for 
the moderate projectile velocities considered
here, the main contribution to the inelastic scattering rate
($\Gamma$) comes from the creation of e-h pairs.
Therefore, from a practical point of view, it is justified 
to consider $\Gamma$ as equivalent to 
the 
probability of e-h pair creation, and we will make no distinction
in what follows.
The energies and oscillator 
strengths of the electronic transition of the Xe atom 
were taken from Ref.\cite{osc}.

Fig.\ref{Fig1}
depicts $\Gamma$ for trajectories parallel to the metal surface with
different incident velocities as a 
function of the impact parameter $z_0$ (measured from the jellium edge)
calculated 
using a) the SRM and, b) our model RPA surface response.
The e-h pair creation probability obtained within
the SRM approximation 
is roughly 
three times smaller,
and decays much faster with $z_0$
than the one calculated using the RPA model response. 
These results indicate that the polarizability of the 
copper surface is seriously underestimated in the simplified
SRM model due to the presence of a fictitious infinite barrier
at the surface. 
In fact, 
the induced electronic charge  
becomes strictly zero beyond the jellium 
edge. This is in contrast with the model RPA response, which peaks at 
the surface and penetrates
some atomic units into the vacuum (see Fig.\ref{Fig3}), 
allowing for higher scattering rates for larger impact parameters.
Our data also point out that, 
in order to obtain a reliable
description of the inelastic scattering at low and intermediate
velocities, it is necessary to explicitly include  
a relatively accurate description of the electronic structure 
in the vicinity of the Fermi level. This is specially true
for the surface band, as will become clearer below.

In Fig.\ref{Fig2} 
we show the scattering rate for both SRM and RPA 
response functions as a function of the velocity of the incident atom, 
for two different impact parameters close to the surface.
At very low $v$, no creation processes are 
possible because the initial energy must at least equal
the energy of the first allowed transition in the incoming atom ($\sim$
9 eV for Xe). 
When $v$ increases, this threshold can be 
overcome, but there is a simultaneous requirement of small 
energy and relatively large momentum transfer to the target that
severely reduces the phase space of the allowed 
electronic excitations
in the substrate.
Increasing $v$ the e-h pair creation 
probability increases until it reaches a maximum 
for $v$ around 2 a.u.. The position of this maximum is 
only slightly dependent on the
impact parameter. By further increasing $v$, $\Gamma$ starts to 
decrease since,
as the energy transfer gets larger, the 
low $q_{\|}$ region available in phase space 
becomes smaller, and the surface response tends to decrease
for large values of $q_{\|}$.
Finally, in the limit of very high $v$ (not reached here) only 
the plasmon contribution becomes important.

The surface state in Cu(111) crosses the Fermi level, so it is a partially
occupied band. This, in addition to the larger weight of this state in 
the surface region, makes reasonable the expectation that a large fraction
of the electrons excited by incoming Xe atoms will be originated from
inter and mainly intraband transitions in this band.
We will study the effect 
of the surface band in the following.
Fig.\ref{Fig3} shows the imaginary part of the
RPA response function for our slab model 
$Im\chi^{RPA}(q_{\|},\omega;z,z^{\prime}=z)$
for $q_{\|}$=0.3 a.u., $\omega$=1 eV, as
a function of $z$ (coordinate normal to the surface), 
with and without
the surface state contribution.
The imaginary part of
the RPA response function gives the number 
of available electronic transitions in the substrate.
While the removal 
of the surface state does not affect the response in
the bulk region, the
number of excitations (e-h pairs in this energy range)
generated at the surface decreases dramatically.

Fig.\ref{Fig4}  shows $\Gamma$ versus the impact 
parameter with and without the surface state, 
for an incident velocity of 2 a.u. Close to the surface, 
the result 
without the surface state is reduced to a $50\%$ 
of what we get with all the states.  
This reduction, however, gets somewhat 
lower as the atom is farther away from the surface.

 In summary, we have reported calculations of the 
inelastic scattering rate of 
neutral atoms with moderate velocities by metal surfaces. 
We have restricted here to the long-range van der Waals term. 
Results for
other contributions to the total inelastic rate 
will be presented elsewhere. 
We have found that the introduction of a more
realistic model of the surface, which explicitly includes 
the electronic structure of the surface near the Fermi energy
and where the response 
function is calculated within RPA approximation, increases 
the scattering rate given by the semiclassical SRM model by
approximately a factor of three. 
These differences are enhanced at low 
and intermediate velocities. By removing
the contribution of the surface state in the 
calculation of the RPA response function, 
we have seen that this band
is the main source electronic excitations in the 
surface region. we conclude then that surface bands
have to be correctly described 
in order to achieve a complete
description of surface-atom scattering events.

\centerline{\bf{Acknowledgments}}
The authors gratefully acknowledge P.M. Echenique, 
J. Garc\'{\i}a de Abajo and A. Garc\'{\i}a-Lekue
for useful discussions. This work has been partially 
funded by the University 
of the Basque Country under Grant UPV/EHU (9/UPV 00206.215-13639/2001) and 
Spanish MCyT under Grant (MAT2001-0946). 
D.S.P acknowledges support from Spanish CSIC and MCyT under the 
"Ram\'on y Cajal" programme.

\begin{figure}
\caption{Inelastic scattering rate for Xe atoms moving parallel to a 
Cu(111) surface as a function
of the impact parameter. The incident velocity is  
$v$=1 a.u. (dot-dashed line),
2 a.u. (dashed) and v=3 a.u. (solid). a) results for the SRM 
surface response 
function, and b) for a model RPA response.
}
\label{Fig1}
\end{figure}

\begin{figure}
\caption{Inelastic scattering rate for Xe atoms moving parallel
to a Cu(111) surface as a function of the incident velocity.
The impact parameter $z_0$ is 1 a.u. (solid lines) and 
2 a.u. (dashed). a) results for the SRM 
surface response
function, and b) for a model RPA response.}
\label{Fig2}
\end{figure}

\begin{figure}

\caption{
Imaginary part of our model RPA response function 
for $q_{\|}$=0.3 a.u. and $\omega$=1 eV, as
a function of $z$ (coordinate normal to the surface), 
for $z=z^{\prime}$ with (solid line), and without (dashed line)
the surface state contribution.}
\label{Fig3}
\end{figure}

\begin{figure}
\caption{Inelastic scattering rate for Xe atoms moving parallel to a
Cu(111) surface as a function
of the impact parameter. The model RPA 
surface response function was used with (solid lines) 
and, without (dashed)
the contribution of the surface state.}
\label{Fig4}
\end{figure}

\end {document}
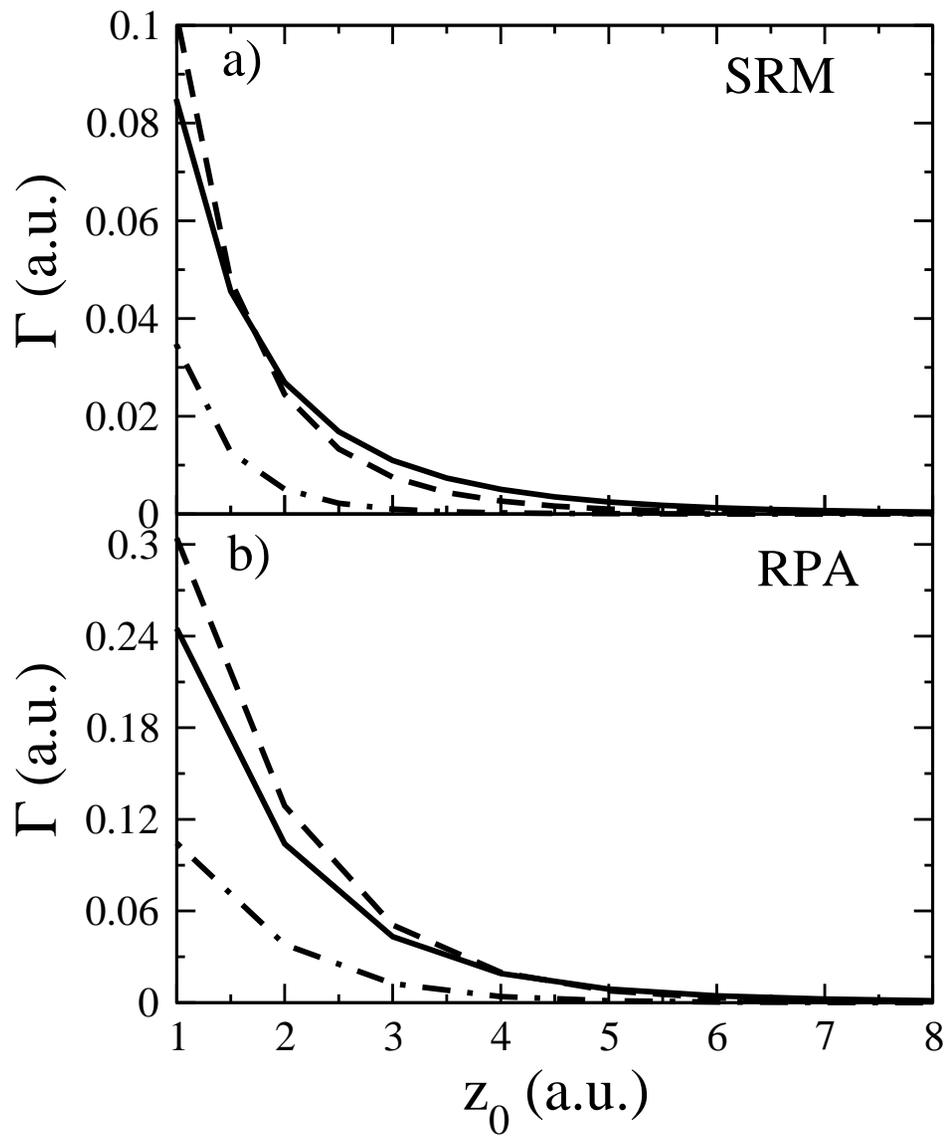
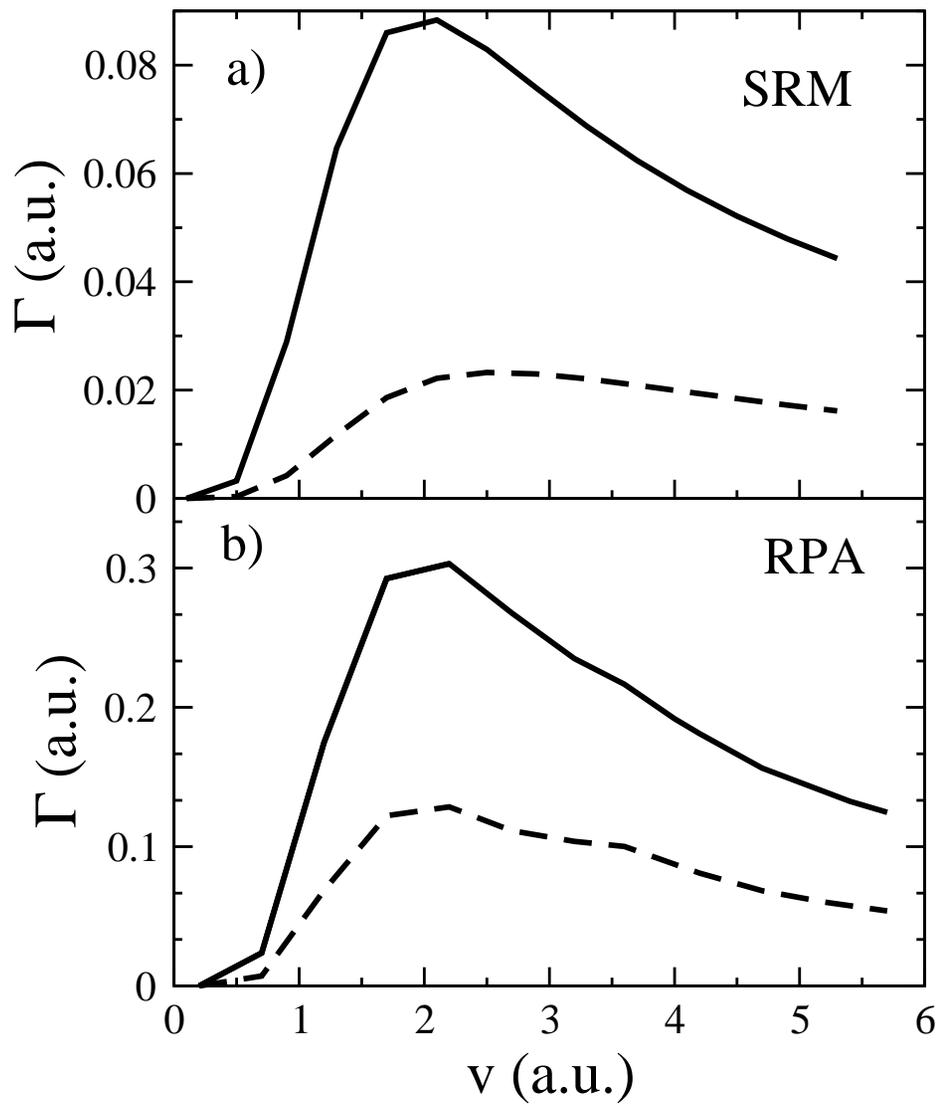
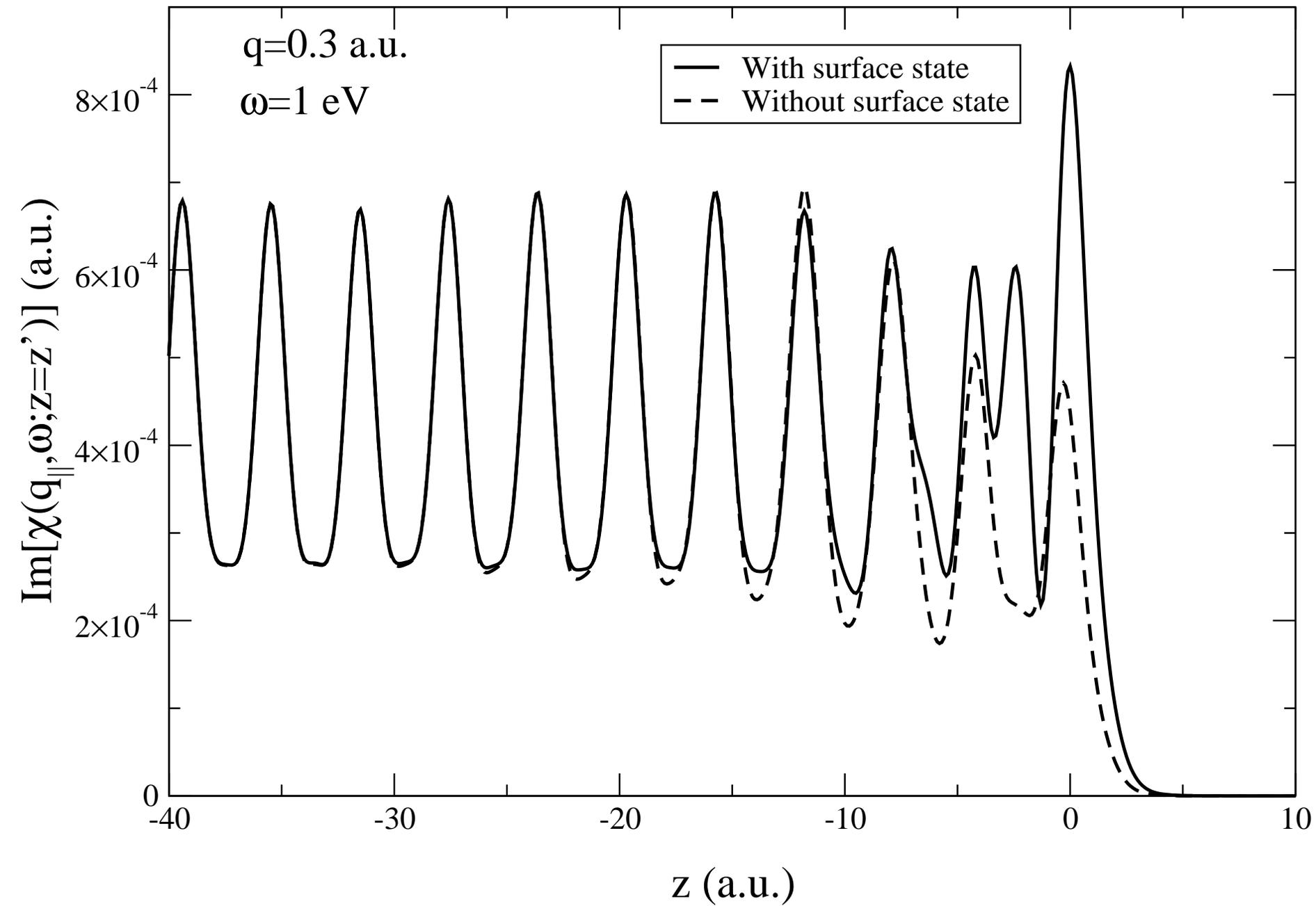
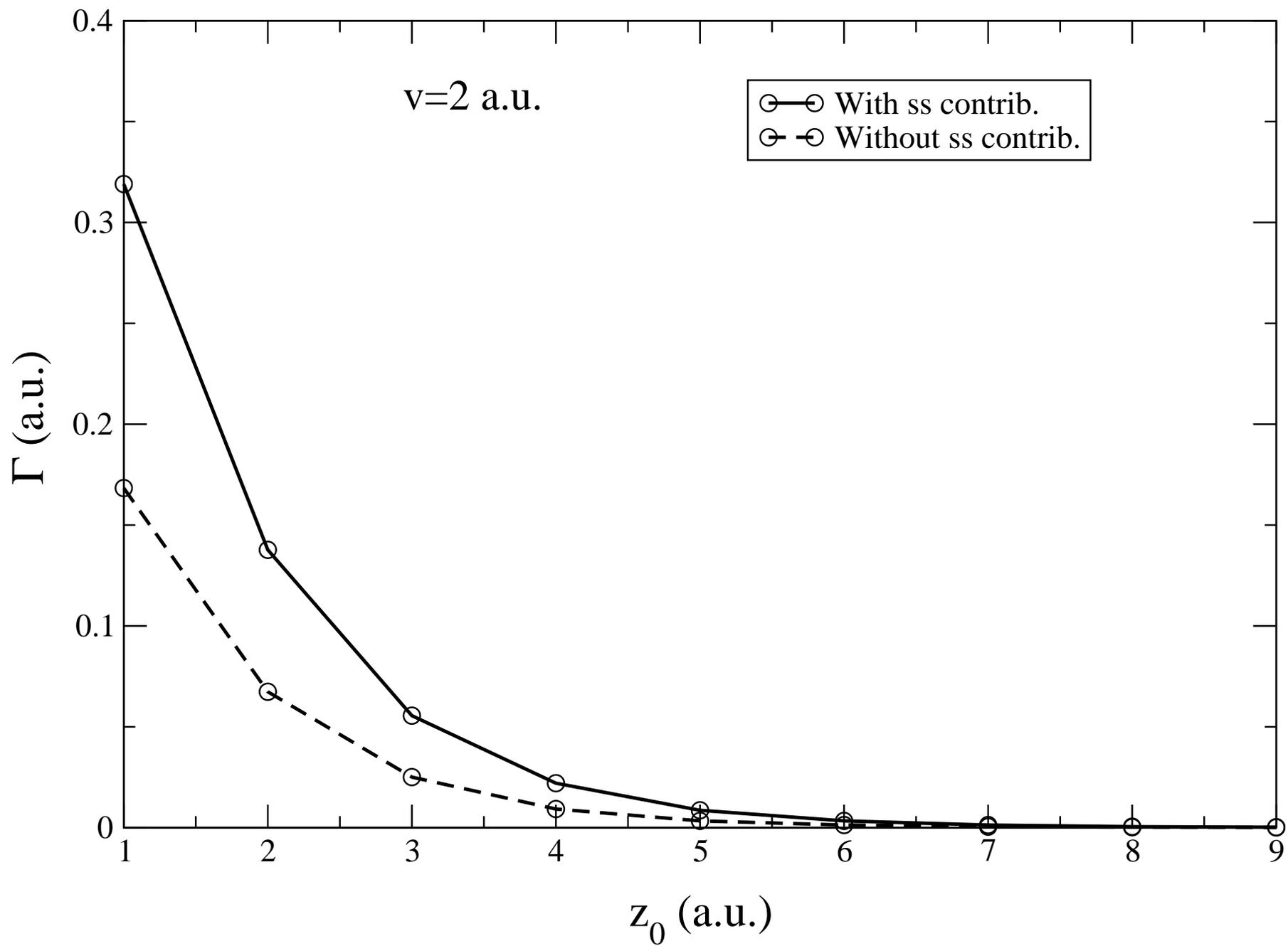
